\documentclass[conference]{IEEEtran}
\IEEEoverridecommandlockouts
\usepackage{cite}
\usepackage{amsmath,amssymb,amsfonts}
\usepackage{algorithmic}
\usepackage{graphicx}
\usepackage{textcomp}
\usepackage{authblk}
\usepackage{todonotes}
 \usepackage[caption=false,subrefformat=parens, labelformat=parens, font=normalsize,labelfont=sf,textfont=sf,margin=20pt]{subfig}

\usepackage{url}

\usepackage{listings}
\def\BibTeX{{\rm B\kern-.05em{\sc i\kern-.025em b}\kern-.08em
    T\kern-.1667em\lower.7ex\hbox{E}\kern-.125emX}}
\begin{document}

\title{Visualizing Design Erosion: \\How Big Balls of Mud are Made}
\author[1]{David Baum}
\author[2]{Jens Dietrich}
\author[3]{Craig Anslow}
\author[1]{Richard M\"uller}
\affil[1]{Leipzig University, Germany \authorcr Email: {\tt \{baum, rmueller\}@wifa.uni-leipzig.de}}
\affil[2]{Massey University, New Zealand \authorcr Email: {\tt J.B.Dietrich@massey.ac.nz}} 
\affil[3]{Victoria University of Wellington, New Zealand \authorcr Email: {\tt craig@ecs.vuw.ac.nz}} 

\maketitle

\begin{abstract}
	Software systems are not static, they have to undergo frequent changes to stay fit for purpose, and in the process of doing so, their complexity increases. It has been observed that this process often leads to the erosion of the systems design and architecture and with it, the decline of many desirable quality attributes, such as maintainability. This process can be captured in terms of antipatterns - atomic violations of widely accepted design principles.
   	We present a visualisation that exposes the design of evolving Java programs, highlighting instances of selected antipatterns including their emergence and cancerous growth. This visualisation assists software engineers and architects in assessing, tracing and therefore combating design erosion. We evaluated the effectiveness of the visualisation in four case studies with ten participants.
\end{abstract}

\section{Introduction}

Software systems are not static, they have to evolve to stay fit for purpose, and as they do, their complexity increases\cite{lehman1980programs}. This tends to have detrimental effects on their quality\cite{banker1993software,khomh2012exploratory} as the ability to adapt a program to changing requirements becomes more and more constrained by its complexity. Consequently, desirable quality attributes suffer. The end stage of this process many projects reach all too quickly has been dubbed spaghetti code or big ball of mud\cite{foote1997big}, while the process itself is often referred to as system rot. 

While measuring the quality of software design is subjective, there is a large body of research trying to assess this by relating it to properties that can be studied by means of static analysis. 
The idea is to extract a representative model from a software system, and then to measure and query it. One such approach is the study of \textit{antipatterns}\cite{koenig1995patterns} and \textit{smells}\cite{fowler1999refactoring}: patterns consisting of artifacts (such as code packages, classes and functions) and their relationships violating certain design principles. 
There are catalogs of widely established principles that facilitate a systematic study of the subject\cite{wiki-antipatterns,smith2000software,bradbury2009defining}.

The evolution of antipatterns is under-researched. Having better insights into their origins and their growth does have benefits that may help software engineers to maintain large projects, and managers to allocate resources for those tasks. For instance, it is useful to know how a part of the system exhibiting strong coupling between modules came about. Tracing this back to the version of the system when the respective dependencies between modules first emerged, and to the respective people and commit messages and other documentation (not) revealing their objectives at the time, will provide valuable information to make informed decisions about a suitable strategy to respond to those issues. The study of design evolution will reveal when bad design becomes rampant, and can relate this to events like product releases under time pressure, team changes, helping with future planning by more correctly assessing the implications of such events.   

We present a visualisation of software design that focuses on evolution in general, and on the evolution of selected antipatterns in particular. The purpose of this visualisation is to assist software engineers to better understand the emergence of design problems. The usage of the visualisation is demonstrated in a screencast\footnote{https://youtu.be/RBgQnE-ozQQ}. Additionally, a demo including an interactive tutorial is available online~\footnote{https://home.uni-leipzig.de/svis/getaviz-antipattern/demo.html}.
We first review related work, followed by a discussion of the visualisation metaphor used and various implementation issues. The evaluation is presented in chapter 4, followed by a brief conclusion.

\section{Related Work}

\paragraph{Antipatterns, Design and Evolution}

The discussion of elements of poor design can be traced back to the early seminal work on software design. 
We chose to study circular dependencies and subtype knowledge (STK) as they directly relate to violations of widely accepted principles of object-oriented design, namely the Acyclic Dependencies Principle\cite{martin2000design} and the Dependency Inversion Principle\cite{riel1996object}. 
What is more, they have precise definitions that facilitate formalisation and therefore the implementation of tools to detect those patterns, and there are algorithms that can be used to detect them that scale well even for large programs.
Circular dependencies was first discussed by Parnas who suggested to keep dependencies between modules \textit{loop free}\cite{parnas1979designing}. 

Empirical studies on larger corpora of real-world programs started in the early 2000s and revealed that surprisingly, antipatterns are prevalent\cite{QualitasCorpus2010}. This was first discovered for circular dependencies\cite{melton2007empirical}, and later confirmed to apply to other antipatterns as well\cite{dietrich2010barriers}. 
Antipatterns can be detected by means of static analysis before a system is deployed. 
The main issue here is the use of dynamic programming language features that create dependencies that may not be visible when the static analysis models are built. This area is generally under-researched, and we must assume that the models used only under-approximate the behaviour of the actual program. In particular, dependency graphs may not contain all edges showing actual program dependencies.

\paragraph{Evolution Visualisation} 

Assessing software quality and improving refactoring decisions are core tasks of software visualisations.
Many visualisations have been developed to support these tasks, e.g. by visualising the systems structure, call graphs, and dependency graphs\cite{Hawes2015, Anslow2013}.
Dependency graphs are usually visualised as node-link-diagrams, enriched with further information\cite{Pinzger2008, Dietrich2008}.
Since these visualisations do not convey any evolutionary information they do not provide any insight about the emergence and evolution of the dependencies. 

There exist many approaches to visualise software evolution\cite{Teyseyre2009}.
Most visualisations want to provide an improved understanding of the development activities by visualising structural changes, e.g. by using added and removed lines as metrics\cite{Schilbach2017,Lanza2005b, Dong2008, Hindle2007, Neu2011, Telea2005, Vernier2016} or by providing highly aggregated information\cite{Sultanow2017,Novais2011,Lanza2005}.
Our use case requires the visualisation of the structural evolution of the system and the antipattern instances at the same time. 
We are not aware of any evolution visualisation that supports this.
There exist evolution visualisations of call graphs\cite{Beyer2006, Fischer2005, Khaloo2017}. However, they do not provide any structural information.



\section{Visualisation Metaphor and Implementation}

\paragraph{Dependency Graph Construction}

The conceptual model of the visualisation presented here is based on a dependency graph extracted from Java bytecode. The graph extraction is based on an ASM-based bytecode analysis\cite{bruneton2002asm}. JDG\cite{jdg} is used to visit the bytecode instructions extracted by ASM and create a directed labelled graph, using data structures from the JUNG network library\cite{o2005analysis}. 
JDG processes bytecode in a two pass process: first, all types are collected and added to the dependency graph to be constructed as vertices. 
In the second pass, JDG tracks all occurrences of types in a particular class file and records them as dependency edges in the graph being constructed, classifying 
them as \textit{extends}, \textit{implements} or \textit{uses} relationships. This corresponds largely with compile time dependencies, although there are some subtle differences. In particular, the Java compiler inlines constants, which leads to a slight under-reporting of compile time dependencies in our model. 

Once a dependency graph has been constructed, it can be queried for antipattern instances. To detect circular dependencies, we use an implementation of Tarjan's algorithm\cite{tarjan1972depth}. 	
To detect STK instances, we use the guery motif engine\cite{dietrich2012scalable}. The extraction pipeline is very similar to the pipeline used by the Massey Architecture Explorer\cite{xplrarc}.

Finally, we rank the vertices representing types in the dependency graph according to their severity by assigning a value between 0 (least severe) and 1 (most severe). For STK, 1 is assigned to the supertype, and 0 to the subtype. Intermediate vertices in the dependency chain connecting the supertype with the subtype are assigned a value of 0.66 if they are abstract, and 0.33 otherwise. Intermediate vertices in the subtype chain from the subtype back to the supertype are all assigned a value of 0.0. 
In the following we will call these values STK rank. While the actual numerical values used here are arbitrary, they reflect the intention of this antipattern as it relates to violations of the dependency inversion principle\cite{martin1996dependency}.

To rank the severity of types in circular dependencies, we use the (min-max-normalized) betweenness centrality\cite{freeman1977set}, computed using Brandes' fast algorithm\cite{brandes2001faster}. The intention here is that it is particularly critical for a class to be in an antipattern if it has more responsibility within the program topology. This is similar to the approach suggested by Martin\cite{martin1995object}, however, by using betweenness centrality over just assessing the relative out-degree we do not only consider the localised impact of dependencies.

\begin{figure}[t]
\includegraphics[width=0.48\textwidth]{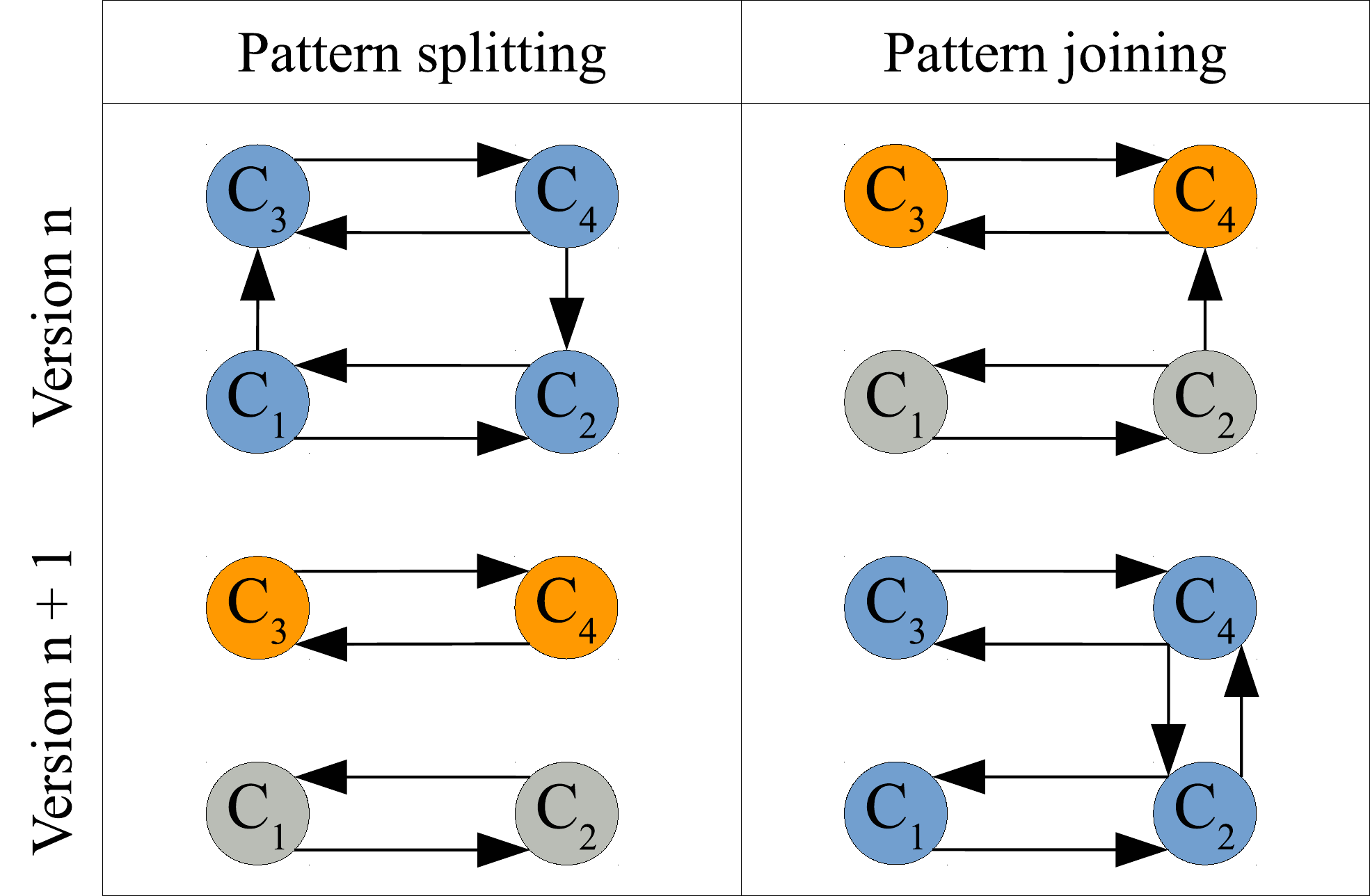}
\caption{Example of one antipattern splitting into two independent antipattern (left) and two independent antipattern merging into one antipattern (right) }
\label{fig:tracing}
\end{figure}

To trace the evolution of an antipattern over multiple versions, it has to be determined whether its occurrence in a version is the result of the evolution of an antipattern in the predecessor version, or a new, different antipattern instance. We define this to be the case if the instance in the successor version has at least 50\% of the types of the antipattern instance in the predecessor version or the other way around. This implies that an antipattern can be split into multiple independent antipattern instances as a system evolves. It is also possible that two independent antipattern instances merge and create a joint antipattern (cf.~Fig.~\ref{fig:tracing}).

\paragraph{Visualisation Design}

\begin{figure}[t]
\centering
   \subfloat[Straight edges]{%
    \includegraphics[width=0.24\textwidth]{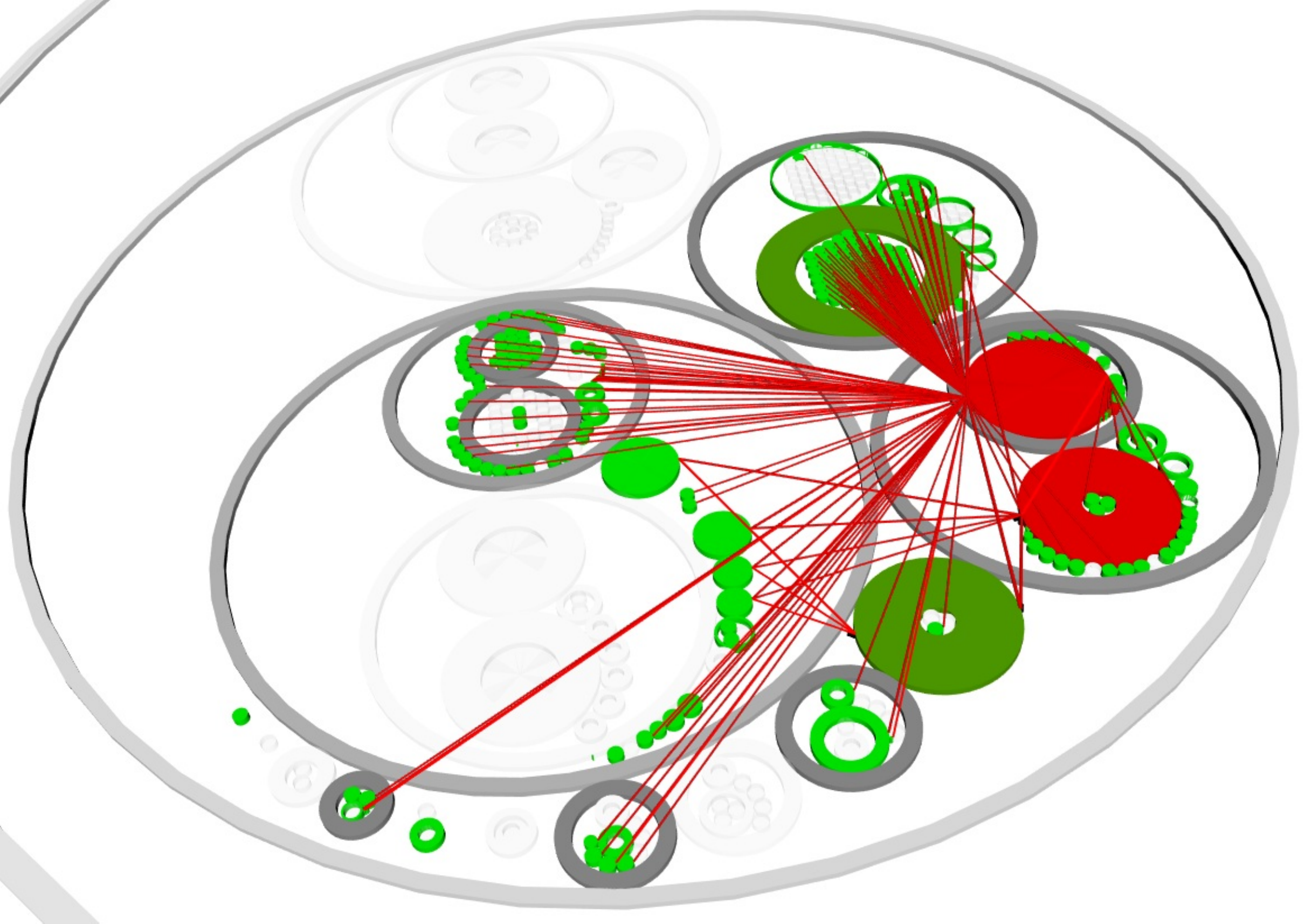}
  }
  \subfloat[Force-directed edge bundling]{%
    \includegraphics[width=0.24\textwidth]{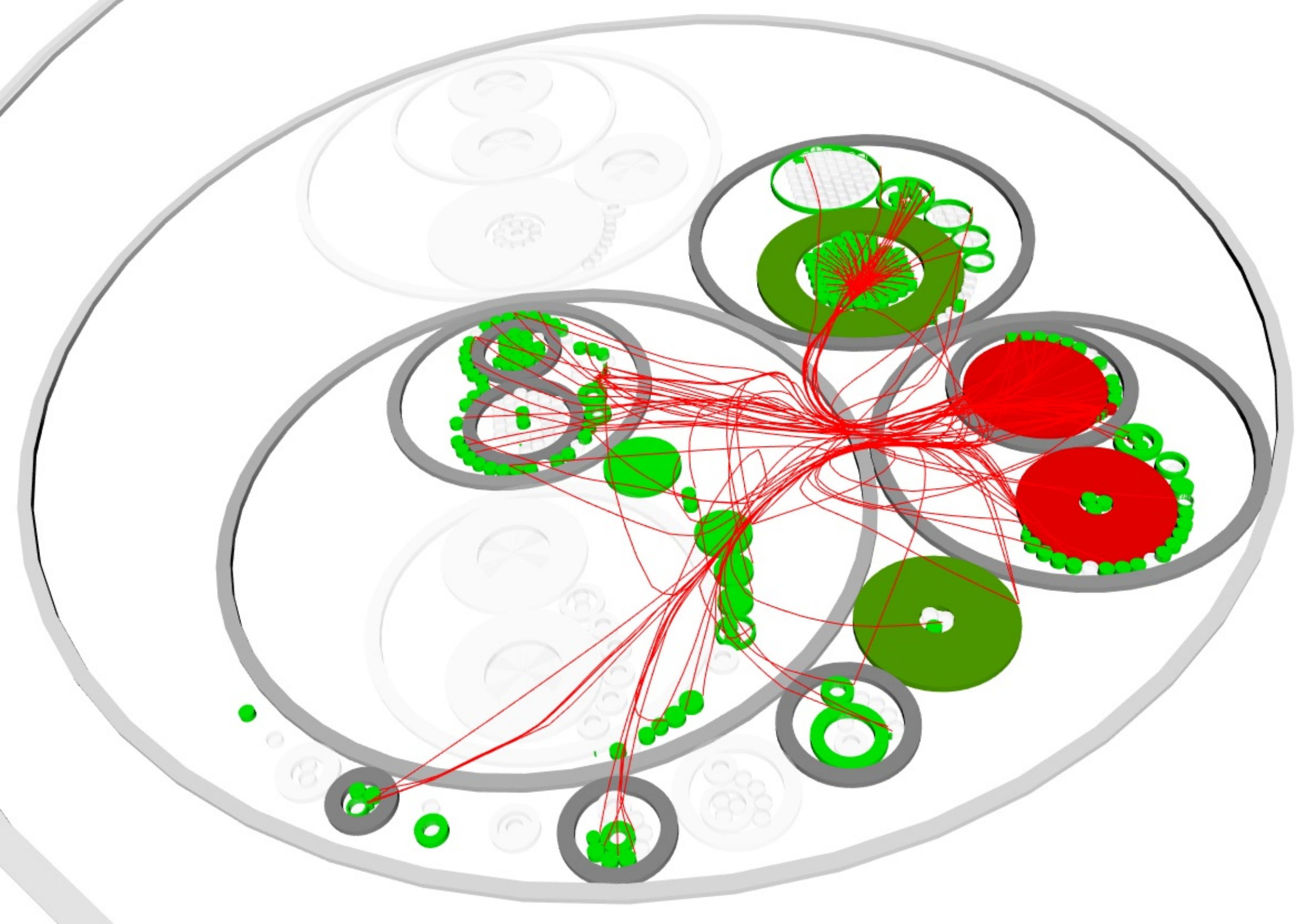}
  }
\caption{Visualisation of the largest cyclic dependency of antlr, including dependencies between 239 classes}
\label{fig:component}
\end{figure}

\begin{figure}[t!]
\includegraphics[width=0.48\textwidth]{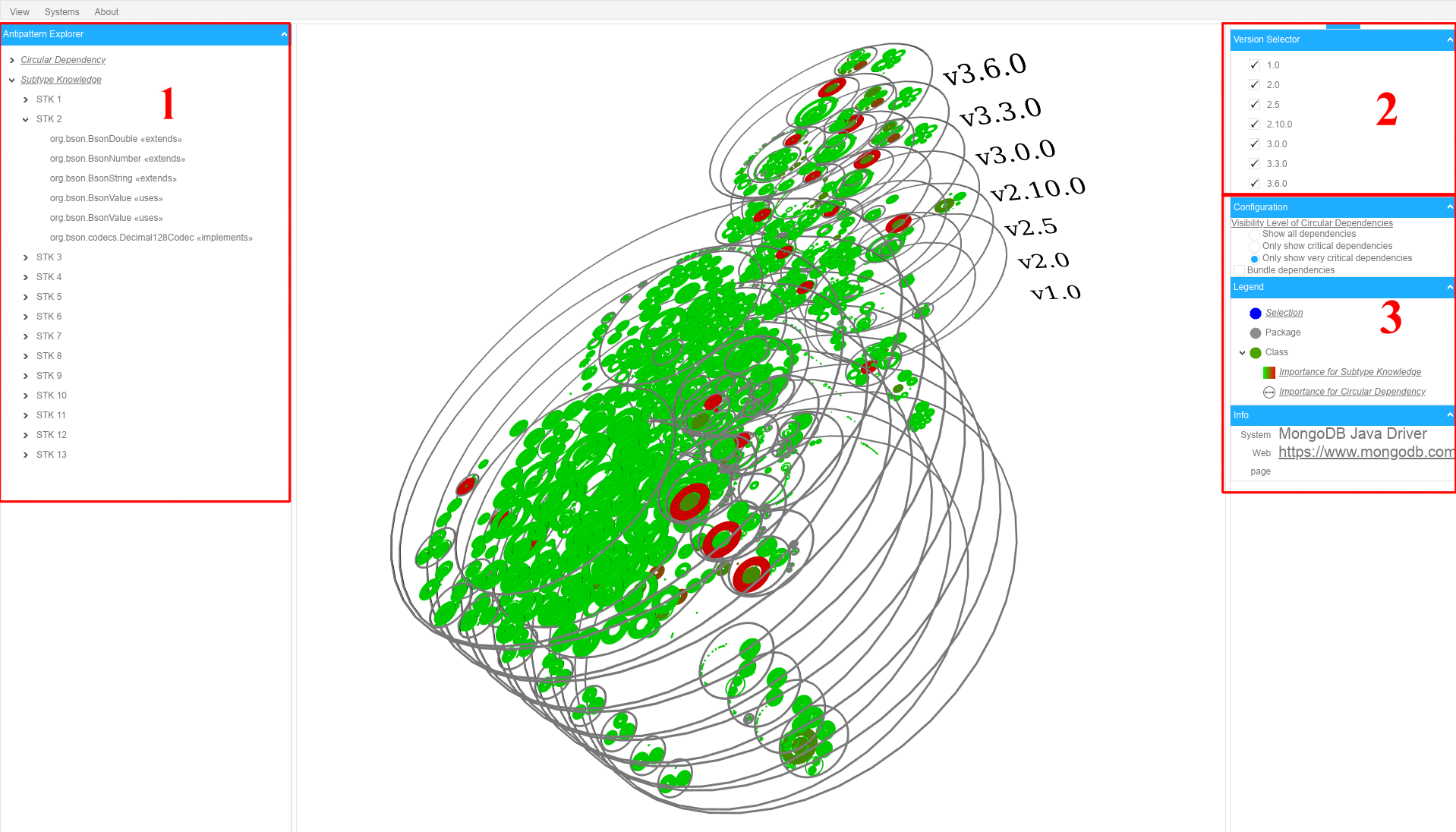}
\caption{Screenshot of Getaviz showing MongoDB Java Driver. 1) Antipattern Explorer 2) Version Selector 3) Legend and Configuration}
\end{figure}

Getaviz~\footnote{https://github.com/softvis-research/Getaviz} is an open source toolkit for the designing and generating software visualisations\cite{Baum2017}.
Getaviz includes the automatic generation of visualisations for several visualisation metaphors. Further, it comes with a  highly configurable browser-based user interface (UI) for viewing and interacting with a visualisation. Currently, the UI only supports X3DOM\cite{x3dom} as rendering platform.
Getaviz can be easily expanded to support new visualisation metaphors and interaction components. Hence, we used Getaviz as starting point and customised the application to fit our requirements for antipattern.

To fit the presented use case a visualisation with many degrees of freedom is necessary, so the structural evolution as well as the antipattern evolution can be visualised at the same time. 
We chose the two-dimensional Recursive Disk (RD) metaphor for structure visualization\cite{Mueller2015b} and enriched it with information about evolution and design erosion.
%
%
%
%
%
%
For every class and package circular disks are used. 
Their area is estimated using the normalised betweenness centrality.
The STK rank is visualised by using a colour scale ranging from green ($0$) to red ($1$). The disks are nested according to the package hierarchy, following a similar presentation in mainstream development tools like IDEs.
Since the developer will work with the visualisation and the IDE at the same time we believe it is important to align both presentations to make it easier to locate entities of interest.


The \textit{Antipattern Explorer} lists all detected antipattern instances in a side bar.
By selecting an instance, all dependencies between the corresponding entities are visualised through red connectors. 
The line thickness reflects the importance of the dependency, so the most critical dependency can be seen on the first glance.
To increase readability the user can choose between straight edges and a forced based edge bundling (cf. Fig.~\ref{fig:component}).
All entities that do not belong to the selected antipattern are greyed out, so the developer can focus on the relevant elements.


Multiple versions can be visualised by piling up the two-dimensional disk visualisations along the $z$-axis which leads to a three-dimensional visualisation.
The $x$ and $y$ coordinates of a disk are stable so that different versions of the same classes are exactly above each other, which makes it easier to visual track classes across different versions. 
The disks are positioned in a helical layout.
This leads to empty space, but reduces occlusion and increases readability. 
Through the \textit{Version Selector} the user can hide uninteresting versions, e.g. minor versions

Alternatively, multiple versions could have been visualised one after the other through animation. 
However, the user needs to remember all classes and relations to spot changes. 
This is error-prone and time-consuming \cite{Schilbach2017}.

Small multiples are a good choice for visualisations that can be viewed at a glance. Software visualisations are large and require navigation. 
Since navigating in multiple visualisations simultaneously can be troublesome, we preferred the three-dimensional representation. 
However, both alternatives are reasonable and have pros and cons.

.




\section{Evaluation}

We investigated the effectiveness of the visualisation based on four case studies. We configured Getaviz to visualise multiple versions of \textit{antlr, JavaMail, MongoDB Java Driver} and \textit{Undertow}. 
The systems were chosen to cover different sized project ranging from 300 classes (JavaMail) to 1,500 classes (Undertow) and includes standard software and research prototypes.

As expected, we found many ``big balls of mud'' across all systems.
Every system contained circular dependencies with several dozen classes.
However, the visualisations did not show how they grew from a small antipattern of only a few classes to an antipattern with over 200 classes. 
In almost every case, antipattern appeared out of nowhere in a version and stayed unchanged in newer versions. We have not found antipattern instances that decreased over time or got dissolved completely. If an instance disappeared, then because the corresponding classes have been removed in the new version. 
This already demonstrates the validity of the visualisation since we had not come to this insight without it and it indicates that developers are not aware of these antipattern instances or do not know how to resolve them.

\begin{figure}[t]
\includegraphics[width=0.46\textwidth]{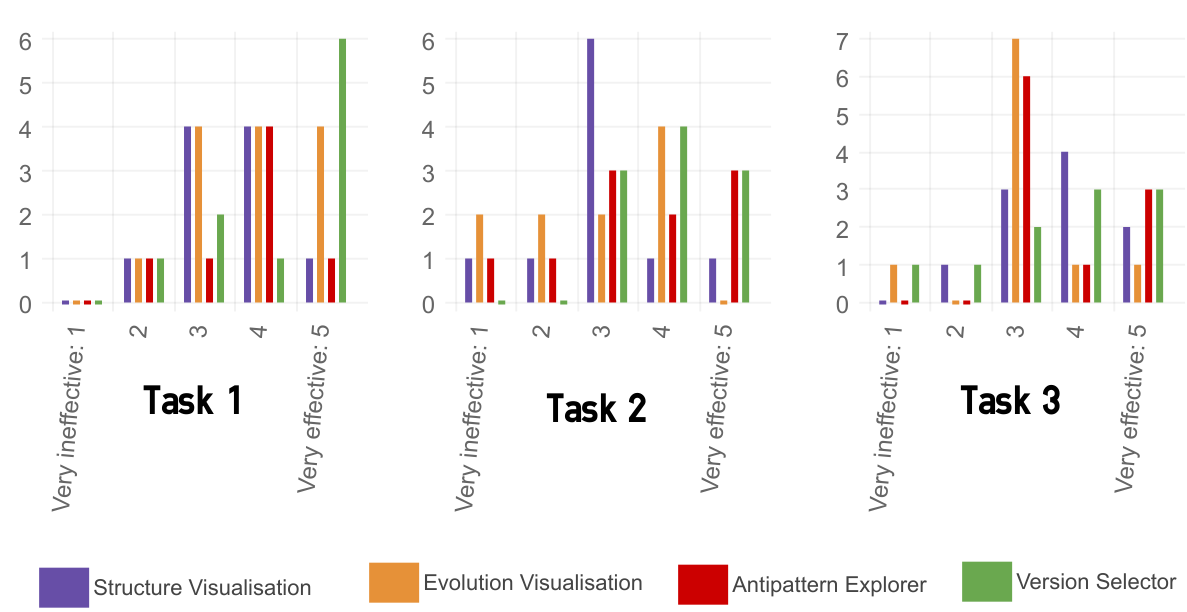}
\caption{Overview over effectiveness rating on Likert skale}
\label{fig:evaluation}
\end{figure}

We invited ten participants (9 male, 1 female) to explore Getaviz. They were not paid and freely opted to participate in the study. All of them have multi-year experience in software development and assess their skills as at least average. 
First, they conducted an interactive tutorial to get familiar with  the visualisation. The evaluation included three comprehension tasks. After each task we asked the participants to rate the effectiveness of different parts of the visualisation (cf. Fig.~\ref{fig:evaluation}). We used a 5-point Likert scale, where $1$ is \textit{very ineffective} and $5$ is \textit{very effective}. Additionally, the participants were asked which aspects they found (in)effective and if they have further suggestions to improve the visualisation. 

\subsection*{Task 1: Which version reduced the quality of the system the most?}

To solve this task, the participants had to compare the design erosion for every version. The visualisation of the structure and the antipattern explorer were rated as slightly effective, the version selector and the representation of multiple versions in parallel were rated more effective. The participants stated that the visualisation can be explored in an intuitive way and comparing versions is easy. They used the version selector to show only one or two versions at the same time.

\subsection*{Task 2: Which packages are part of the original Circular Dependency Component 1?}

To solve this task, the participants had to identify the first appearance of the circular dependency and gather the corresponding packages names through hovering over them to see the tooltip. The visualisation of the structure and the version layering were evaluated as slightly ineffective. The antipattern explorer was rated as slightly effective. The participants liked especially the visualisation of the dependencies between the classes. The version selector was again the most effective part. In order to solve this task the participants had to navigate several times within the visualisation. This was the most challenging part of the task. Participants sometimes lost track of the elements of interest or needed several attempts to move the visualisation to the desired position.

\subsection*{Task 3: With which class would you start refactoring?}

This task refers to the most recent version only. Hence, the visualisation of multiple versions is superfluous. Therefore, it was rated as neither effective nor ineffective. The participants used the version selector and rated it as effective. However, some participants stated that it is time consuming to hide every uninteresting version individually and it would be more convenient to switch the displayed version with one click.
The structure visualisation and the antipattern explorer were rated as more effective. The participants stated that problematic classes are easy to detect, but that the visualisation is too cluttered on the one side and lacks further information to answer the question studiously on the other side. 

\section{Discussion}

Almost all refactoring decisions of the participants are reasonable. 
We are satisfied with these initial results, although there is significant potential for improvement and some design choices should be reconsidered. In the following we would like to discuss some problems identified by the study and how they could be improved in future. Some ideas arose from reviewing the participants answers, some were suggested from the participants directly.

\paragraph{Technical limitations}
The complaints of the participants about navigation concerns the sometimes confusing behaviour of X3DOM and not the actual visualisation. For instance, zoom via scroll wheel works opposed to the usual behaviour. 
The visualisation makes extensive use of transparency. 
The transparency support in X3DOM is defective. 
In some cases transparent elements are displayed opaque so that other elements are occluded.
The version layering would work much better with a better support for transparency. To overcome these restrictions we will switch from X3DOM to Mozilla's A-Frame, which provides superior transparency handling and navigation capabilities.

\paragraph{Colour Mapping}

The disk colour depicts only one isolated quality aspect. This is misleading as users might interpret it as an overall quality aspect even if the legend states otherwise. In Task 3, most participants chose classes represented by red vertices. This is not necessarily wrong, but might indicate that they made a decision mainly based on this colour. 
To overcome this issue we may prefer a more neutral colour palette\cite{Stone2002}.

\paragraph{Version layering}

The version layering was useful for some tasks, but can lead to complicated navigation issues and cluttered visualisations. As already stated, the situation could be improved through migrating to A-Frame. Nevertheless, the participants perceived an information overload and reduced the visible versions to one or two. Still, tracing antipatterns through different versions was quite effective and was expressly praised by two participants. Therefore, it is probably the best solution to support the current version layering as well as small multiples in order to support more tasks effectively.

\paragraph{Supported tasks}

The visualisation has to cover more tasks and quality measures to be a comprehensive visual analytics tool for assessing the design erosion of large and complex software systems. 
For example, only one antipattern instance can be highlighted currently. For assessing the overall quality of a version it would be better to highlight all instances at the same time and use different colours to distinguish them. Further, Getaviz should support more antipatterns. Once the antipatterns are detected by static analysis tools they can be integrated in the visualisation easily.


\paragraph{Scalability}

Scalability is a known issue for large software visualisations, especially when multiple versions are depicted. We are capable to visualise systems with up to 500,000 LOC and about ten versions at the same time. However, the visualisation can consist of many more versions if they are loaded on demand.

\paragraph{Threats to Validity}

We conducted only a preliminary study with ten participants.
An extensive evaluation is necessary once the biggest issues revealed have been solved. 
The largest system of the evaluation has about 1,500 classes. The visualisation might become more confusing on larger systems due to more edge crossings and a higher number of involved classes in general. Hence, the effectiveness of the visualisation has to be evaluated for large systems in a controlled manner.
The participants rated the effectiveness of the visualisation without a direct comparison to different solution approaches, e.g. doing the tasks directly in an IDE or using conventional visualisations. 

\section{Conclusion}

We demonstrated that Getaviz is an easy to adopt framework for the visualisation of program evolution. We were able to visualise the erosion of systems design and architecture exemplary for two antipatterns, cyclic dependencies and STK. 
The validation with end users indicates that the tool has potential to assist software engineers in gaining a better understanding of design erosion, and to use this understanding for corrective refactoring. 
Findings from our evaluation revealed significant potential for improvement, to be addressed in future work.

\bibliographystyle{plain}
\bibliography{references.bib,david.bib}

\end{document}